\documentclass[conference]{IEEEtran}
\IEEEoverridecommandlockouts
\usepackage{cite}
\usepackage{amsmath,amssymb,amsfonts}
\usepackage{algorithmic}
\usepackage{graphicx}
\usepackage{textcomp}
\usepackage{xcolor}
\usepackage{subcaption}
\def\BibTeX{{\rm B\kern-.05em{\sc i\kern-.025em b}\kern-.08em
    T\kern-.1667em\lower.7ex\hbox{E}\kern-.125emX}}
\begin{document}

\title{Who is Authentic Speaker?\\
}

\author{\IEEEauthorblockN{Qiang Huang}
\IEEEauthorblockA{\textit{School of Computer Science} \\
\textit{University of Sunderland}\\
Sunderland, UK \\
qiang.huang@sunderland.ac.uk}

}

\maketitle

\begin{abstract}
Voice conversion (VC) using deep learning technologies can
now generate high quality one-to-many voices and thus has
been used in some practical application fields, such as entertainment
and healthcare. However, voice conversion can pose
potential social issues when manipulated voices are employed
for deceptive purposes. Moreover, it is a big challenge to
find who are real speakers from the converted voices as the
acoustic characteristics of source speakers are changed greatly.
In this paper we attempt to explore
the feasibility of identifying authentic speakers from converted
voices. This study is conducted with the assumption that certain
information from the source speakers persists, even when their
voices undergo conversion into different target voices. Therefore
our experiments are geared towards recognising the source
speakers given the converted voices, which are generated
by using FragmentVC on the randomly paired utterances from
source and target speakers. To improve the robustness
against converted voices, our recognition model is constructed by
using hierarchical vector of locally aggregated descriptors (VLAD)
in deep neural networks. 
The authentic speaker recognition system is mainly
tested in two aspects, including the impact of quality of converted
voices and the variations of VLAD. The dataset
used in this work is VCTK corpus, where source and
target speakers are randomly paired. The results obtained on the
converted utterances show promising performances in recognising
authentic speakers from converted voices.
\end{abstract}


\section{Introduction}
Speech synthesis is one of most attractive research topics in speech processing.
Belonging to this field, voice conversion aims to transform one voice from a source 
speaker to the sound like another person's voice
without changing the linguistic content \cite{Childers19889}.  
A typical voice conversion system is generally input with
utterance pairs from the source and target speakers.
The speech waveform from source speakers are converted
into a compact representation linking to the phonetic information, 
while the acoustic features from target speakers' voices are
extracted for mapping. The mapping or
conversion function is trained on these aligned mapping features. In
the conversion phase, after computing the mapping features from
a new source speaker utterance, the features are converted using
the trained conversion function. The speech features are computed
from the converted features which are then used to synthesize the
converted utterance waveform \cite{MOHAMMADI201765, Walczyna2023, Wu2014}.

In the last ten years, the use of deep neural networks has significantly
boosted speech synthesis technologies and enables voice conversion to be used
in some fields, such as entertainment industry \cite{Mukhneri2020, Huang2023_VC, Chen2019SingingVC} 
and healthcare \cite{Rioja2023, Raman2021}. However, using voice conversion
can also cause some potential ethical and social issues.
For example, utilizing a converted voice resembling that of a well-known actor or singer 
without proper consent can result in legal complications.
For healthcare, confidential spoken information from patients
also needs be protected from being misused.
Besides these, there also exist potential risks that converted voices could
be used for deceit, which enables people to believe the fake information conveyed by the converted
voices that people could be familiar with. 
Due to these reasons, concerns have been raised regarding privacy and authentication.
Therefore, preventing the incorrect use of
one’s voice with voice conversion technologies becomes more and
more important \cite{Cai2023}.

To tackle these potential issues, there have been some studies in detecting synthesized voices
in order to distinguish real human voices from synthesized ones \cite{Mari2022,borzi2022synthetic,mo2022multi}.
These studies mainly rely on the use of deep neural networks to build a binary classifier to
identify whether input voices are true or not. However, some recent studies show that 
using generative adversarial networks (GANs) can mitigate the detection ability \cite{Doan2023, Li2024}.
Unlike those previous work in detection, our work aims to go deeper to find who is the authentic
source speaker given converted voices. Although there have been some conversion algorithms, such as encoder-decoder
models \cite{chen2021} and GAN based models \cite{Kaneko2018, Hirokazu2018}, these methods still can not
completely eliminate speaker-dependent features.

\begin{figure*}[t]\label{fig:framework}
  \centering
  \includegraphics[clip, trim=3.5cm 6cm 4.1cm 7cm, scale=0.6]{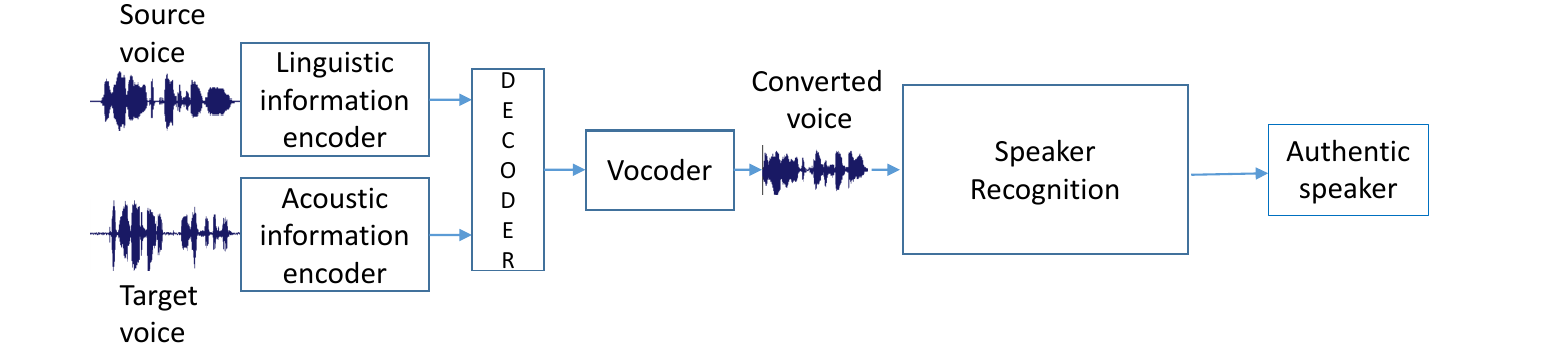}
  \caption{Architecture of authentic speaker recognition system, mainly including 
  voice conversion and speaker recognition}\label{fig:asr2}
\end{figure*} 
 
In this work, we aim to construct a speaker recogniser to identify authentic source
speakers given converted voices, transformed by using an encoder-decoder model.
As aforementioned, converted voices contain 
the information from the source speaker,
however their acoustic characteristics are typically significantly compressed 
when transitioning from waveforms to vectors.
Moreover target speaker's information is dominant in the converted voices
and interferes the search for the information from source speaker.
This poses a significant challenge for authentic speaker recognition using converted
voices.
Although some methods \cite{Cai2018ExploringTE} 
for speaker recognition have been developed,
most of them were designed to handle conventional voices, 
even when they are affected by noise. 
To effectively learn source speaker's features from converted voices,
a dictionary-based NetVlad embedded within a ResNet trunk architecture  
is used in this work to process variable-length of transformed utterances
by aggregating features across time \cite{Xie2019}.

For the details of this work, they can be found in the next sections.
This paper is organised as follows: Section 2 presents the theoretical framework
of this paper by introducing voice conversion and the construction of speaker
recogniser; Section 3 depicts the data to be used, experimental setup, and
the metrics used for evaluation;
Experimental results are analysed in detail in Section 4, and finally conclusion
is drawn and future work is discussed in Section 5.


\section{Theoretical Framework}\label{sec:framework}
Our system consists of two parts, voice conversion (VC)
and authentic speaker recognition. Voice conversion aims to convert
the input utterances $\textbf{U}_{s}$ from source speakers $\textbf{\textit{S}}$ to
the utterances $\textbf{U}_{vc}$ having similar voice style to the given target utterances $\textbf{U}_{t}$ from
target speakers $\textbf{\textit{Ts}}$. 
Speaker recognition is to identify the authentic source speaker $S_{a}$ from all
possible source speakers $\textbf{\textit{S}}$.  
Figure 1 shows the architecture of the system to be developed 
for authentic speaker recognition.

\subsection{Voice Conversion}\label{subsec:vc}
As aforementioned, the mechanism of voice conversion used in this work
is encoder-decoder model. Its encoder needs to learn phonetic
information from source speakers and extract acoustic from target
speaker. The decoder aims to combine the extracted information using decoder
before the information goes to vocoder.   

Wav2Vec 2.0 \cite{Alexei2020} is used to extract
the features relevant to linguistic information of $U_s$. 
It consists of multiple convolutional layers, takes
as input source utterance $U_s$, and outputs speech representations
$z_1, ..., z_T$ over T time steps, followed by a transformer $g: Z \rightarrow C$ 
used to map the information of entire audio sequence to representation vectors
$c_1, ..., c_T$ \cite{Alexei2020, Jacob2019}.
To extract acoustic features of target utterances and combine with the source 
information, a U-Net \cite{Olaf2015} like structure is used.
Extracting the acoustic information from target utterances mainly relies on stacked 1D-convolutional
layers, whose output is sent to a decoder. 

The aim of decoder is to reconstruct spectrogram-level information by fusing the information from
both source and target speakers. It consists of extractors and smoothers.
The extractors are based on the latent phonetic structure of the
source speaker utterance, while
the smoother considers the high correlation among adjacent features in speech by using
self-attention \cite{FragmentVC}.
Finally, a WaveRNN-based speaker-independent
vocoder is used to convert the reconstructed 
log mel-spectrograms to waveforms by speech synthesis. Its details are
outlined in \cite{Jamie2019}.

\subsection{Authentic Speaker Recognition using VLAD}\label{subsec:asr}
The architecture of speaker recogniser consists of three blocks:
feature extraction, feature aggregation, and classification.
In general, the basic feature extraction is to convert 2D spectrogram into 1D bottleneck vector
before it goes to classifier. In this work, ResNet \cite{He2016deep} 
is used as backbone architecture.

Feature aggregation aims to map variable-length inputs to a fixed-length template
descriptors, which have larger similarities of templates of the same subject than that of
different subjects \cite{Zhong2018}. In our work, Vector of Locally
Aggregated Descriptors (VLAD)\cite{Jegou2010} is used to conduct
feature aggregation.

VLAD originally was used for learning image feature descriptor.
It aims to group the local feature descriptors of all the images into 
a couple of clusters, and computes the algebraic sum of the residue vectors between 
each cluster centroid and the descriptors of a specific image belonging to this cluster. 
In this work, the input feature vector of VLAD are learned from the network extracting
audio spectrogram.
For $N$ $D$-dimensional input feature $x_i$ and a chosen number of clusters K, 
VLAD produces $v$ according to the following equation:

\begin{equation}
    v_{j,k} = \sum_{i=1}^N x_i(j) - c_k(j)
\end{equation}
where $x_i(j)$ and $c_k$ respectively denote the $j$th component of
the input feature $x$ considered and its corresponding cluster $c_k$. 
The vector $v$ is subsequently performed L2 normalization using
$v = v/||v||_2$.

\begin{figure}[tbh]
  \includegraphics[clip, trim=0cm 0cm 1cm 0.0cm, scale=0.36]{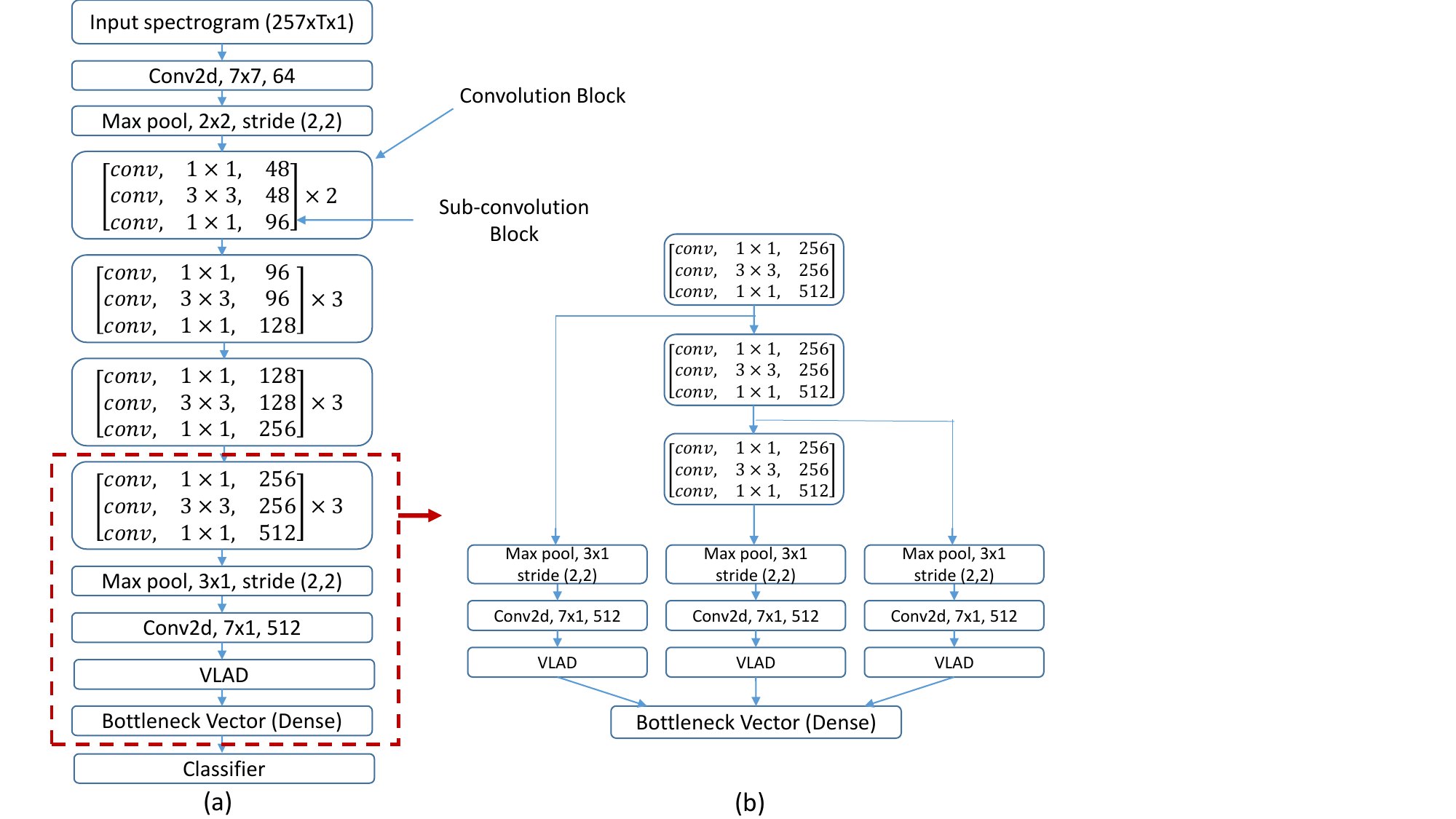}
  \caption{Hierarchical VLAD for authentic speaker recognition: 
  (a) Architecture of thin ResNet34 followed by VLAD and classifier; 
  (b)proposed structure by linking the output of each sub-convolution 
  block in the last convolutional block with an individual VLAD layer.}\label{fig:hvlad}
  
\end{figure}
   
\subsection{Hierarchical VLAD}\label{subsec:cvlad}

In general a deeper architecture can benefit the refinement of some important 
information to meet targets. However, some subtle information relevant to target could be 
ignored although residue structure is taken into account to reduce the potential impact. 
Moreover VLAD works as a kind of information quantization to some extends, the possibility of
losing useful information could be also increased. In the proposed architecture, multiple VLADs
are used to connect with different convolution layers instead of using only one VLAD layer before
information goes to classifier.

Figure \ref{fig:hvlad}(a) shows the architecture of a speaker recognition baseline system,
including encoder and classifier.   
The used backbone of its encoder is RESNET34, containing multiple convolution blocks. 
These convolution blocks consisting of two or three sub-convolution blocks, each of them having
three 2D convolutional layers. A VLAD layer is inserted
behind Maxpooling layer followed by a fully connected (FC) layer used to generate bottleneck vector. 
Figure \ref{fig:hvlad}(b) shows the proposed hierarchical structure by expanding
the last convolution block in the baseline system. 
In the proposed structure, the output of each sub-block will be connected to a VLAD layer,
and then goes to a shared FC layer. The use of a shared FC layer aims to
reduce computation cost and improve the robustness of compact vector representation for classification.

\section{Data}
Voice conversion systems trained with parallel data are not ideal in 
real world because parallel data including the same contents from source and
target speakers is hard to obtain.
So the VC used in our work uses only non-parallel voice data by randomly selecting utterances
from target speakers.

In this work, CSTR VCTK \cite{Junichi2019} was used to convert voices from source speakers to target speakers.
This CSTR VCTK Corpus was originally aimed for speech synthesis using average voice models trained on multiple speakers and speaker adaptation technologies. This Corpus includes speech data from 110 English speakers 
(two speakers, p280 and p315, have issues with recording) with various accents. 
Each speaker reads out about 400 sentences, which were selected from a variety of resources, including
newspaper, rainbow passage, and elicitation paragraph. The rainbow passage and elicitation paragraph
are the same for all speakers. All speech data was recorded using an identical recording setup and were converted into 
16 bits, downsampled to 48 kHz, and manually end-pointed \cite{Junichi2019}. 

To conduct voice conversion, we need to pair source and target utterances.  
In our experiments, 100 utterances were randomly selected without repetition from each speaker 
and are used as source utterances. Their corresponding target utterances were also randomly selected from any other speakers, 
excluding the original speakers. So there are 10800 (100$\times$108) converted utterances in total.


\section{Experimental Setup}
In this paper, our experiments consists of two parts, voice conversion and authentic speaker recognition.
Voice conversion aims to generate target voices, which will be used to identify the authentic speakers 
by speaker recognition.

Voice conversion is conducted by using FragmentVC \cite{FragmentVC}. 
It can implement any-to-any voice conversion by learning the latent phonetic 
structure of the utterance from the source speaker by using Wav2Vec \cite{Alexei2020}
and extracting the spectral features (log mel-spectrograms) of the utterances
from the target speakers. By aligning the hidden structures
of the two different feature spaces with a two-stage training process,
FragmentVC is able to extract fine-grained voice fragments from the
target speaker utterance(s) and fuse them into the desired utterance,
all based on the attention mechanism of Transformer as verified with
analysis on attention maps, and is accomplished end-to-end \cite{FragmentVC}.
Although this voice conversion system is trained with reconstruction loss
only without any disentanglement considerations between content and speaker
information. Moreover, this conversion system does not require parallel data,
which is like some real deceitful scenarios when using voice conversion.

To extract the phonetic features of source speaker's voices, a pretrained model was used in Wav2Vec 2.0 
to extract 768-dimensional speech representations,
without finetuning model weights. The 768-dimensional features
are then converted to 512-dimension by two linear layers with
ReLU activation, to be used as the input to the decoder.

For authentic speaker recognition, the converted utterances were split into two parts,
9000 utterances for training and the rest 1800 utterances for testing. Recognition accuracy was
used as an evaluation metric in this work.
To conduct speaker recognition, a 2.5-second segment was randomly extracted
from each converted utterance. Spectrograms are generated by using a 512 point
FFT on short fragments obtained by using a 25ms sliding window with a 10ms hop size.
The spectrogram is normalised by subtracting the
mean and dividing by the standard deviation. 
Although the use of a dynamic learning rate in \cite{Xie2019} showed better performances,
this case was not found in our experiments. So Adam optimiser with a fixed learning rate of 0.0001
was used in this work.

\section{Result and Analysis}
Our experiments start with measuring the similarity between the converted voice
and source speaker's voice. A marking criterion, Mean Opinion Score (MOS) \cite{wikiMOS},
was designed. It includes 5 marks, ranging from 1 to 5. 1 indicates absolutely different and 5 means
absolutely same. 
Some examiners listened to an authentic utterance from the source speaker
and a converted utterance, and then gave a mark from 1 to 5 
to show how confident they thought the two utterances were from
the same speaker.

\begin{figure}[t]
  \centering
  \includegraphics[clip, trim=3.0cm 10cm 3.5cm 10cm, scale=0.6]{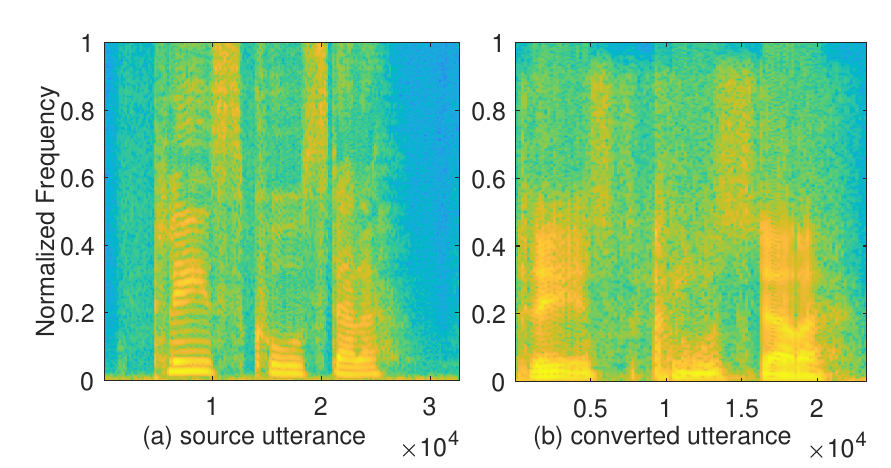}
  \caption{Spectrogram of source utterance and converted utterance.}\label{fig:spec}
\end{figure}


\begin{table}[th]
\small
\caption{Subjective evaluation of similarity between source and converted utterances}\label{tab:mos}
\centering
 \begin{tabular}{|c| c |} 
 \hline
\#target utterances  & MOS   \\ [0.1ex] 
  \hline
 1 & 1.39$\pm$0.18 \\ \hline
 2 & 1.33$\pm$0.15 \\ \hline
 3 & 1.30$\pm$0.16\\

 \hline
\end{tabular}
\end{table}

Table \ref{tab:mos} shows MOS obtained by averaging the marks from 10 examiners 
across 30 randomly selected converted utterances. 
It is clear that the voices from source speakers and the converted voices are
quite different. 
In this table, MOS was also compared when the voices were converted when using
different number of target utterances. 
However, there are only slight differences among these MOS under the three
conditions. These results indicate identifying authentic source speaker is very difficult
by using a subjective evaluation. 

Although quality evaluation is not presented in this paper due to space limitation,
an example is shown in Fig \ref{fig:spec}. It showcases the spectrograms of source utterance and converted
utterance. Some similar phonetic information can be found in the two
examples, however the quality of converted utterance is relatively poor
compared to the source utterance.
This is because only three target utterances were used for acoustic extraction

\begin{table}[th]
\small
\caption{Comparison of recognition performances obtained by using HVLAD and baseline methods}\label{tab:acc}
\centering
 \begin{tabular}{|c| c|} 
 \hline
\#target  Model  &  Accuracy on test (\%)   \\ [0.1ex] 
  \hline
  Baseline1  & 13.50 $\pm$ 0.28 \\ 
  Baseline2  & 14.16 $\pm$ 0.27 \\
  Baseline3  & 14.50 $\pm$ 0.25 \\
  HVLAD  &   15.38   $\pm$ 0.25\\

 \hline
\end{tabular}
\end{table}

Table \ref{tab:acc} shows comparisons of recognition accuracy obtained on the test data
by using hierarchical VLAD and three baseline methods:

\begin{description}
\item[\textbf{Baseline1:}] $~~~~~$
uses the architecture of Fig \ref{fig:hvlad}(a), but replaces the VLAD layer with a Flatten layer
\item[\textbf{Baseline2:}]  $~~~~~$
uses the architecture of Fig \ref{fig:hvlad}(a) only
\item[\textbf{Baseline3:}]  $~~~~~$
uses the architecture of Fig \ref{fig:hvlad}(a), but replaces the last block with Fig \ref{fig:hvlad}(b)
where the VLAD layer is replaced by a Flatten layer
\item[\textbf{HVLAD:}]  $~~~~~$
uses the architecture of Fig \ref{fig:hvlad}(a), but replaces the last block with Fig \ref{fig:hvlad}(b) only
\end{description}

The results obtained on the test data show that Baseline1
yields the worst recognition accuracy.
Although the use of Baseline2 outperforms Baseline1 due to the use of VLAD, its architecture
lacks the ability to collect target-relevant information from previous layers and
sub-blocks. So Baseline3 shows slight better performance than Baseline2 since the hierarchical
structure probably helps learn useful information. After combining the two factors,
VLAD and hierarchical structure, HVLAD does best than the three baseline methods.     
In these experiments, VLAD in Baseline2 and HVLAD has 64 clusters. The converted
voices were generated by using one target utterance.

Unlike conventional speaker recognition where the acoustic characteristics of speakers
to be identified are generally dominant even if they are corrupted by noise, 
authentic speaker recognition focuses on subtle information of source speakers
from converted voices. This requires the used neural network
is sensitive to learn subtle features relevant to source speakers. The use of VLAD can mitigate
possible interferences from target speakers, and the hierarchical structure
can retrieve subtle information from different layers. It is probably the reason why 
HVLAD can work better than others.

\begin{figure}[tbh]
  \centering
  \includegraphics[clip, trim=3.5cm 8cm 4.1cm 9cm, scale=0.5]{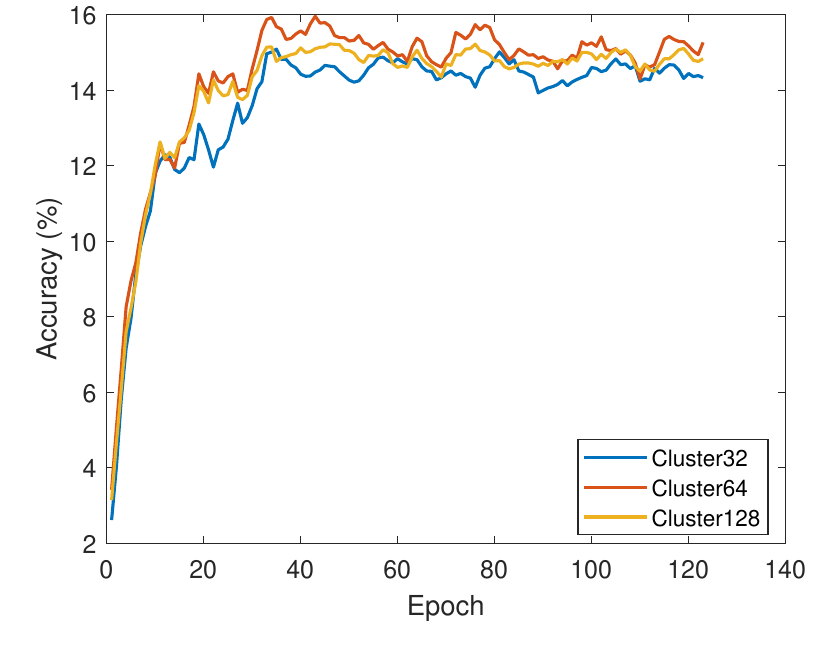}
  \caption{Authentic speaker recognition accuracy when varying the number of VLAD clusters.}\label{fig:asr1}
\end{figure}

To further explore the effectiveness of HVLAD, different number of VLAD clusters are set for
evaluation. Fig. \ref{fig:asr1} showcases the recognition accuracy on the test data when
the number of cluster is set to 32, 64, and 128, respectively.
When the number of cluster is 32, recognition performance 
is instability.
When the number of cluster is increased to 64, 
the recognition accuracy can reach 16$\%$. 
However increasing the number of Vlad cluster to 128 does not further 
improve recognition accuracy, although the performance is more stable than
previous two.

\begin{figure}[h!]
  \centering
  \includegraphics[clip, trim=3.5cm 8cm 4.1cm 9cm, scale=0.5]{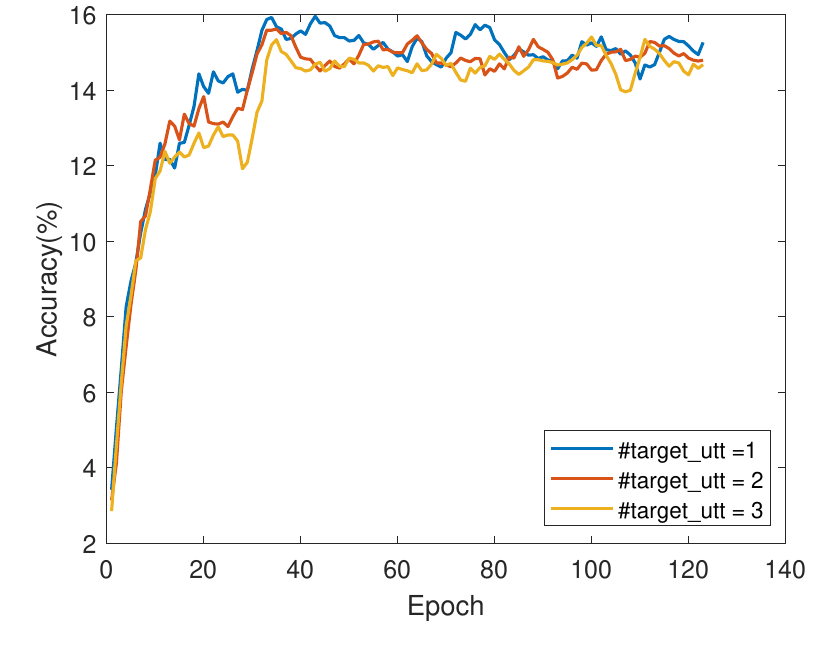}
  \caption{Authentic speaker recognition accuracy when using different number of target utterances}\label{fig:asr2}
\end{figure}

To evaluate how the number of target utterances impacts recognition accuracy, further
experiments were conducted. Fig \ref{fig:asr2} shows  
the accuracies obtained when the number of target utterances is 1, 2, and 3, respectively. In this experiment,
the number of Vlad cluster was set to 64.
The curves shown in this figure undergo smoothing done by averaging the accuracy values of five neighbouring data points.
Although some differences can be found among the three curves, they are not significant. 
This could be related to three reasons. The first reason is the number of target utterances used for
voice conversion is small. The acoustic features extracted from target speakers might be insufficient
to generate voices similar to target speakers. The second reason could be related to the case that
source utterance and target utterances are not in parallel, which might have impact on the quality
of the converted voices. The third reason could be relevant to the existence of acoustic features
from source speakers. Although the quality of the converted utterances is relatively poor and the similarity
between the converted utterance and the source utterance is low, the acoustic characteristics of
source speaker might not be completely eliminated even the number of target utterance is increased from 1 to 3.

\begin{table}[th]
\footnotesize
\caption{Top-1 and Top-5 authentic speaker recognition accuracy}\label{tab:top1_5}
\centering
 \begin{tabular}{|c c c c|} 
 \hline
 \multicolumn{4}{|c|}{\textbf{Impact of number of target utts on accuracy ($\%$)}}\\
 \multicolumn{4}{|c|}{($\#$Vlad cluster $=$ 64)}\\
 \hline
  & 1 target utt & 2 target utts & 3 target utts \\ [0.1ex] 
 \hline
 Top-1 & 15.38$\pm$0.25 & 15.00$\pm$0.31 & 14.53$\pm$0.37 \\
 Top-5 & 20.67$\pm$0.21 & 20.50$\pm$0.23 & 20.02$\pm$0.25 \\
 \hline
 \hline
 \multicolumn{4}{|c|}{\textbf{Impact of number of Vlad clusters on accuracy ($\%$)} }\\
 \multicolumn{4}{|c|}{($\#$target utt $=$ 1)}\\
 \hline
  & 32 clusters & 64 clusters & 128 clusters \\ [0.1ex] 
 \hline
 Top-1 & 14.52 $\pm$ 0.33 & 15.38 $\pm$ 0.25 & 14.97 $\pm$0.29 \\
 Top-5 & 19.22 $\pm$ 0.38 & 20.67 $\pm$ 0.21 & 20.05$\pm$0.23 \\
 \hline
\end{tabular}
\end{table}

Table \ref{tab:top1_5} shows top-1 and top-5 authentic speaker recognition
accuracy. 
The upper table discusses the impact of the number of target utterances
on accuracy. The increase of target utterances to three does not show significant
reduction in accuracy, while more authentic speakers can be found from five 
top-ranked candidates. The lower table compares top-1 and top-5 identification performances
when changing the number of VLAD clusters. It is clear that the number of Vlad
clusters has an impact on the accuracy especially when using 32 Vlad clusters in
the recogniser. Moreover, the number of identified authentic speakers within top-5 candidates
is also less than the other two cases configured with 64 and 128 clusters, respectively. 
As the number of phoneme is about 50, the optimal number of VLAD clusters might be somewhat relevant to it.

\section{Conclusion and Future Work}

In this work, we explored the feasibility to recognise authentic source speakers
from converted voices and constructed a novel recogniser by 
combining a hierarchical structure with VLAD. The initial experiments show
that the use of this proposed model can yield better performances than
three baseline methods using only VLAD or the hierarchical structure.
Additional experiments were conducted to evaluate the impacts caused by the number of
VLAD clusters and the number of target utterances were conducted. The obtained
results show that the optimal number of clusters might be relevant to the number of
phonemes. The increase of target utterances could improve the quality
of converted voices, however it will probably lead to further challenges
in authentic speaker recognition.

Our future work will consider using some advanced deep learning technologies, 
such as attention mechanism and the methods used in speech separation, to
learn robust audio features. In addition, we will consider to construct
a benchmark dataset to support the relevant research in this field.

\bibliographystyle{IEEEtran}
\bibliography{mybib}

\begin{thebibliography}{10}
\providecommand{\url}[1]{#1}
\csname url@samestyle\endcsname
\providecommand{\newblock}{\relax}
\providecommand{\bibinfo}[2]{#2}
\providecommand{\BIBentrySTDinterwordspacing}{\spaceskip=0pt\relax}
\providecommand{\BIBentryALTinterwordstretchfactor}{4}
\providecommand{\BIBentryALTinterwordspacing}{\spaceskip=\fontdimen2\font plus
\BIBentryALTinterwordstretchfactor\fontdimen3\font minus
  \fontdimen4\font\relax}
\providecommand{\BIBforeignlanguage}[2]{{%
\expandafter\ifx\csname l@#1\endcsname\relax
\typeout{** WARNING: IEEEtran.bst: No hyphenation pattern has been}%
\typeout{** loaded for the language `#1'. Using the pattern for}%
\typeout{** the default language instead.}%
\else
\language=\csname l@#1\endcsname
\fi
#2}}
\providecommand{\BIBdecl}{\relax}
\BIBdecl

\bibitem{Childers19889}
D.~Childers, K.~Wu, D.~Hicks, and B.~Yegnanarayana, ``Voice conversion,''
  \emph{Speech Communication}, vol.~8, no.~2, pp. 147--158, 1989.

\bibitem{MOHAMMADI201765}
S.~H. Mohammadi and A.~Kain, ``An overview of voice conversion systems,''
  \emph{Speech Communication}, vol.~88, pp. 65--82, 2017.

\bibitem{Walczyna2023}
\BIBentryALTinterwordspacing
T.~Walczyna and Z.~Piotrowski, ``Overview of voice conversion methods based on
  deep learning,'' \emph{Applied Sciences}, 2023. [Online]. Available:
  \url{https://doi.org/10.3390/app13053100}
\BIBentrySTDinterwordspacing

\bibitem{Wu2014}
Z.~Wu and H.~Li, ``Voice conversion versus speaker verification: An overview,''
  \emph{APSIPA Transactions on Signal and Information Processing}, vol.~3, 12
  2014.

\bibitem{Mukhneri2020}
F.~Mukhneri, I.~Wijayanto, and S.~Hadiyoso, ``Voice conversion for dubbing
  using linear predictive coding and hidden markov model,'' \emph{Journal of
  Southwest Jiaotong University}, vol.~55, 01 2020.

\bibitem{Huang2023_VC}
W.-C. Huang, L.~Violeta, S.~Liu, J.~Shi, and T.~Toda, ``The singing voice
  conversion challenge 2023,'' 12 2023, pp. 1--8.

\bibitem{Chen2019SingingVC}
\BIBentryALTinterwordspacing
X.~Chen, W.~Chu, J.~Guo, and N.~Xu, ``Singing voice conversion with
  non-parallel data,'' \emph{2019 IEEE Conference on Multimedia Information
  Processing and Retrieval (MIPR)}, pp. 292--296, 2019. [Online]. Available:
  \url{https://api.semanticscholar.org/CorpusID:73729115}
\BIBentrySTDinterwordspacing

\bibitem{Rioja2023}
\BIBentryALTinterwordspacing
I.~Hernáez-Rioja, J.~A. Gonzalez-Lopez, and H.~Christensen, ``Special issue on
  applications of speech and language technologies in healthcare,''
  \emph{Applied Sciences}, pp. 2--13, 2023. [Online]. Available:
  \url{https://doi.org/10.3390/app1311684}
\BIBentrySTDinterwordspacing

\bibitem{Raman2021}
\BIBentryALTinterwordspacing
S.~Raman, X.~Sarasola, E.~Navas, and I.~Hernaez, ``Enrichment of oesophageal
  speech: Voice conversion with duration–matched synthetic speech as
  target,'' \emph{Applied Sciences}, pp. 2--13, 2021. [Online]. Available:
  \url{https://doi.org/10.3390/app11135940}
\BIBentrySTDinterwordspacing

\bibitem{Cai2023}
D.~Cai, Z.~Cai, and M.~Li, ``Identifying source speakers for voice conversion
  based spoofing attacks on speaker verification systems,'' 2023.

\bibitem{Mari2022}
D.~Mari, F.~Latora, and S.~Milani, ``The sound of silence: Efficiency of first
  digit features in synthetic audio detection,'' in \emph{2022 IEEE
  International Workshop on Information Forensics and Security (WIFS)}, 2022,
  pp. 1--6.

\bibitem{borzi2022synthetic}
S.~Borz{\`\i}, O.~Giudice, F.~Stanco, and D.~Allegra, ``Is synthetic voice
  detection research going into the right direction?'' in \emph{Proceedings of
  the IEEE/CVF Conference on Computer Vision and Pattern Recognition}, 2022,
  pp. 71--80.

\bibitem{mo2022multi}
Y.~Mo and S.~Wang, ``Multi-task learning improves synthetic speech detection,''
  in \emph{ICASSP 2022-2022 IEEE International Conference on Acoustics, Speech
  and Signal Processing (ICASSP)}.\hskip 1em plus 0.5em minus 0.4em\relax IEEE,
  2022, pp. 6392--6396.

\bibitem{Doan2023}
T.~P. Doan, K.~Hong, and S.~Jung, ``Gan discriminator based audio deepfake
  detection,'' in \emph{Proceedings of the 2nd Workshop on Security
  Implications of Deepfakes and Cheapfakes}, 2023, pp. 29–--32.

\bibitem{Li2024}
F.~Li, Y.~Chen, H.~Liu, Z.~Zhao, Y.~Yao, and X.~Liao, ``Vocoder detection of
  spoofing speech based on gan fingerprints and domain generalization,'' pp.
  1--20, 2024.

\bibitem{chen2021}
Y.-H. Chen, D.-Y. Wu, T.-H. Wu, and H.-y. Lee, ``Again-vc: A one-shot voice
  conversion using activation guidance and adaptive instance normalization,''
  in \emph{ICASSP 2021 - 2021 IEEE International Conference on Acoustics,
  Speech and Signal Processing (ICASSP)}, 2021, pp. 5954--5958.

\bibitem{Kaneko2018}
T.~Kaneko and H.~Kameoka, ``Cyclegan-vc: Non-parallel voice conversion using
  cycle-consistent adversarial networks,'' 2018, pp. 2100--2104.

\bibitem{Hirokazu2018}
H.~Kameoka, T.~Kaneko, K.~Tanaka, and N.~Hojo, ``Stargan-vc: Non-parallel
  many-to-many voice conversion using star generative adversarial networks,''
  2018, pp. 266--273.

\bibitem{Cai2018ExploringTE}
W.~Cai, J.~Chen, and M.~Li, ``Exploring the encoding layer and loss function in
  end-to-end speaker and language recognition system,'' \emph{ArXiv}, vol.
  abs/1804.05160, 2018.

\bibitem{Xie2019}
W.~Xie, A.~Nagrani, J.~S. Chung, and A.~Zisserman, ``Utterance-level
  aggregation for speaker recognition in the wild,'' 05 2019, pp. 5791--5795.

\bibitem{Alexei2020}
A.~Baevski, H.~Zhou, A.~Mohamed, and M.~Auli, ``Wav2vec 2.0: a framework for
  self-supervised learning of speech representations,'' 2020, pp.
  12\,449--12\,460.

\bibitem{Jacob2019}
J.~Devlin, M.-W. Chang, K.~Lee, and K.~Toutanova, ``Bert: Pre-training of deep
  bidirectional transformers for language understanding,'' 2019, pp.
  4171--4186.

\bibitem{Olaf2015}
O.~Ronneberger, P.~Fischer, and T.~Brox, ``U-net: Convolutional networks for
  biomedical image segmentation.''\hskip 1em plus 0.5em minus 0.4em\relax
  Springer International Publishing, 2015, pp. 234--241.

\bibitem{FragmentVC}
\BIBentryALTinterwordspacing
Y.~Y. Lin, C.-M. Chien, J.~hao Lin, H.~yi~Lee, and L.-S. Lee, ``Fragmentvc:
  Any-to-any voice conversion by end-to-end extracting and fusing fine-grained
  voice fragments with attention,'' in \emph{IEEE International Conference on
  Acoustics, Speech and Signal Processing (ICASSP)}, 2021, pp. 5939--5943.
  [Online]. Available: \url{https://api.semanticscholar.org/CorpusID:225076127}
\BIBentrySTDinterwordspacing

\bibitem{Jamie2019}
J.~Lorenzo-Trueba, T.~Drugman, J.~Latorre, T.~Merritt, B.~Putrycz,
  R.~Barra-Chicote, A.~Moinet, and V.~Aggarwal, ``Towards achieving robust
  universal neural vocoding,'' 2019, pp. 181--185.

\bibitem{He2016deep}
K.~He, X.~Zhang, S.~Ren, and J.~Sun, ``Deep residual learning for image
  recognition,'' in \emph{Proceedings of the IEEE conference on computer vision
  and pattern recognition}, 2016, pp. 770--778.

\bibitem{Zhong2018}
Y.~Zhong, R.~Arandjelovi{\'{c}}, and A.~Zisserman, ``Ghostvlad for set-based
  face recognition,'' in \emph{Computer Vision -- ACCV 2018}, 2019, pp. 35--50.

\bibitem{Jegou2010}
H.~Jégou, M.~Douze, C.~Schmid, and P.~Pérez, ``Aggregating local descriptors
  into a compact image representation,'' in \emph{IEEE Computer Society
  Conference on Computer Vision and Pattern Recognition}, 2010, pp. 3304--3311.

\bibitem{Junichi2019}
J.~Yamagishi, C.~Veaux, and K.~MacDonald, ``Cstr vctk corpus: English
  multi-speaker corpus for cstr voice cloning toolkit (version 0.92),''
  University of Edinburgh. The Centre for Speech Technology Research (CSTR).

\bibitem{wikiMOS}
\BIBentryALTinterwordspacing
W.~contributors, ``Mean opinion score, the free encyclopedia,'' 2024. [Online].
  Available: \url{https://en.wikipedia.org/wiki/Mean-opinion-score}
\BIBentrySTDinterwordspacing

\end{thebibliography}

\end{document}